\documentclass[10pt, journal]{IEEEtran}
\newcommand{\subparagraph}{}
\usepackage{cite}
\usepackage{amsmath,amssymb,amsfonts}
\usepackage{graphicx}
\usepackage{textcomp}
\usepackage{flushend}
\usepackage[font=footnotesize]{caption}
\usepackage{comment}
\usepackage{physics,amsmath}
\include{pythonlisting}
\usepackage{float}
\usepackage{graphics}
\usepackage{titlesec}
\usepackage{subfig}
\usepackage{graphicx}
\usepackage{tabularx}
\usepackage{siunitx}

\usepackage{xcolor}
\usepackage{titlesec}
\usepackage{subfig}
\usepackage{graphicx}
\usepackage{mathtools}
\usepackage{cite}
\usepackage[font=footnotesize]{caption}
\usepackage{array}
\usepackage[english]{babel}
\usepackage[utf8x]{inputenc}
\usepackage{xcolor}
\usepackage[colorinlistoftodos]{todonotes}

\usepackage{enumitem}

\usepackage{amsmath}
\usepackage{amssymb}
\usepackage{amsfonts}
\usepackage{amstext}
\usepackage{amsthm}
\usepackage{soul}
\usepackage{multicol}
\usepackage{algpseudocode}
\usepackage{tikz}
\usepackage{url}

\titlespacing{\section}{0pt}{2ex}{1ex}
\titlespacing{\subsection}{0pt}{0.5ex}{0.2ex}
\titlespacing{\subsubsection}{0pt}{0.5ex}{0ex}

\newcommand*\darkgreycircled[1]{\tikz[baseline=(char.base)]{
            \node[shape=circle,fill=darkgray,text=white,draw=black,inner sep=0.1pt] (char) {#1};}}

\begin{document}
\title{TimeFloats: Train-in-Memory with Time-Domain Floating-Point Scalar Products}
\author{Maeesha Binte Hashem$^1$, Benjamin Parpillon$^{1,2}$, Divake Kumar$^1$, Dinithi Jayasuriya$^1$, and Amit Ranjan Trivedi$^1$ \\
$^1$ AEON Lab, University of Illinois at Chicago (UIC), Chicago, IL - USA \\
$^2$ Fermi National Accelerator Lab (FNAL), Batavia, IL - USA \\
} 

\maketitle
\begin{abstract}
In this work, we propose ``TimeFloats,'' an efficient train-in-memory architecture that performs 8-bit floating-point scalar product operations in the time domain. While building on the compute-in-memory paradigm's integrated storage and inferential computations, TimeFloats additionally enables floating-point computations, thus facilitating DNN training within the same memory structures. Traditional compute-in-memory approaches with conventional ADCs and DACs face challenges such as higher power consumption and increased design complexity, especially at advanced CMOS nodes. In contrast, TimeFloats leverages time-domain signal processing to avoid conventional domain converters. It operates predominantly with digital building blocks, reducing power consumption and noise sensitivity while enabling high-resolution computations and easier integration with conventional digital circuits. Our simulation results demonstrate an energy efficiency of 22.1 TOPS/W while evaluating the design on 15 nm CMOS technology. 

\end{abstract}

\section{Introduction}
Deep learning has revolutionized the field of artificial intelligence, enabling computers to learn from vast amounts of data and make predictions with unprecedented accuracy. From image and speech recognition to natural language processing and self-driving cars, deep learning has had a dramatic impact on numerous applications and industries. The ability of state-of-the-art deep learning models such as GPT, BigGAN, T5, BERT, and AlphaGo Zero \textit{etc.}, to uncover complex patterns and insights from large datasets has transformed the way various applications such as robotics and healthcare approach decision making. State-of-the-art deep learning models however also require substantial computational resources to train, often utilizing high-powered GPUs or TPUs. Their training results in large energy consumption and high levels of carbon emissions. For example, GPT-3, with over 175 billion parameters, requires several days of training on multiple GPUs, emitting thousands of tons of carbon dioxide equivalent (CO$_2$e), comparable to the lifetime emissions of tens of cars!

Current deep learning models are trained on centralized servers (e.g., cloud computing) and then transmitted to edge devices like smartphones, IoT, or embedded systems for inference. Transmitting large datasets from edge devices to centralized nodes increases carbon emissions due to transmission energy overhead. To ensure sustainable deep learning, therefore, it is essential to develop hardware and methods that support \textbf{edge learning} on local devices, reducing dependence on centralized servers. This is even more critical for real-time applications like autonomous drones and medical devices. Furthermore, edge learning can also enhance efficiency, privacy, and security by keeping data local while offering flexibility and adaptability in scenarios with limited connectivity.

\begin{figure}[t!]
    \centering
     \includegraphics[width=\linewidth]{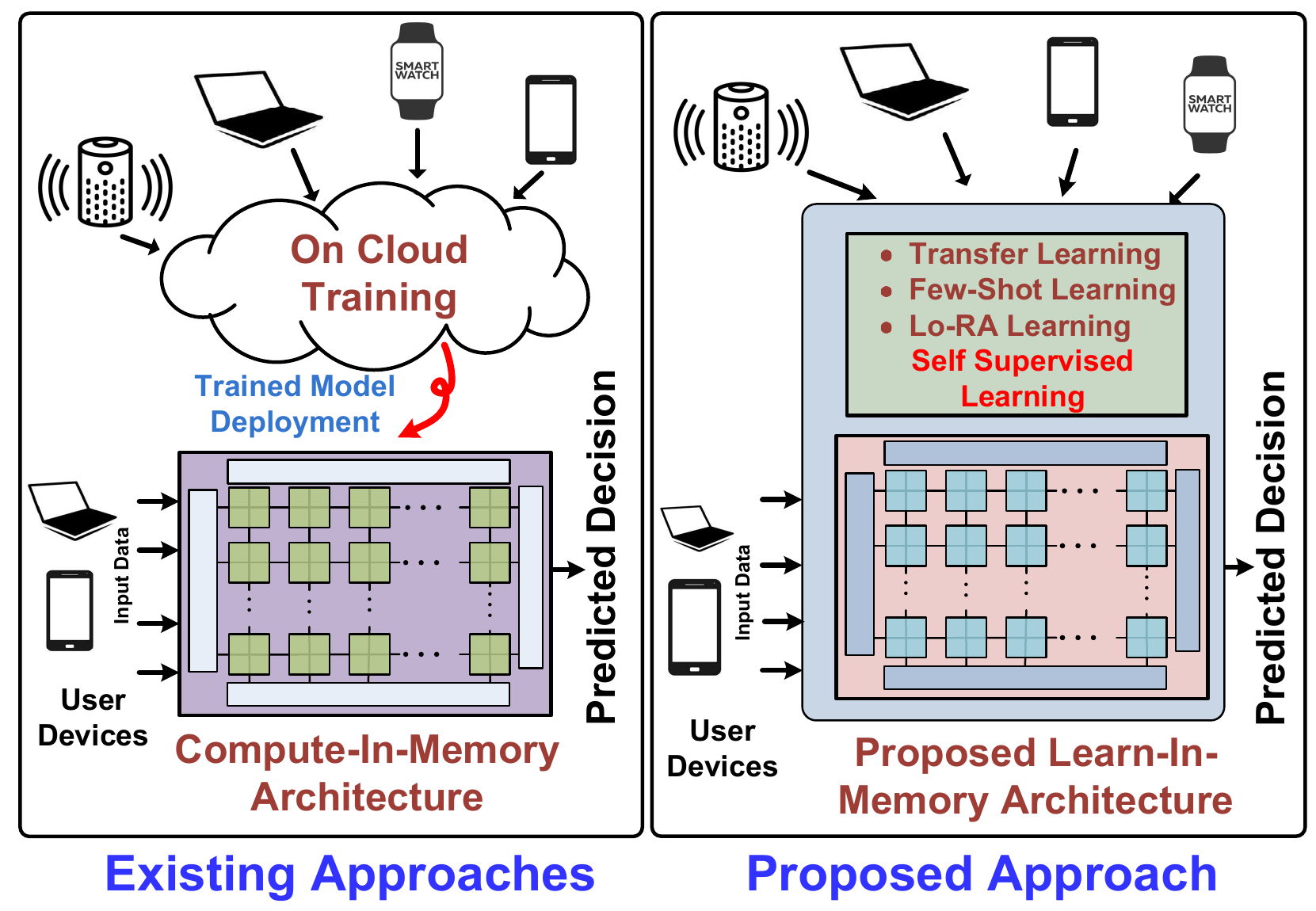}
    \caption{\textbf{Edge Learning with Train-in-Memory:} Current DNN training relies on the cloud due to the data-intensive nature of the training procedures. In contrast, this work presents train-in-memory using time-domain computing capable of performing efficient training directly within memory.}
    \label{fig:Overview}
\end{figure}

Addressing deep learning's training challenges, this paper explores \textit{train-in-memory}, building on the compute-in-memory paradigm, dominant in low-power edge inference. In compute-in-memory, the same memory array handles both weight storage and inference, reducing data movement by collapsing operations into a single module. We extend this concept by enabling storage, inference, and training on the memory array, through a novel architecture called as \textit{TimeFloats}, equipping edge devices with both learning and inference capabilities. While fixed precision suffices for inference, training requires floating-point arithmetic for handling gradients \cite{gupta2015deep}. For floating-point in-memory processing, \textit{TimeFloats} provides the following fundamental advantages:

\begin{itemize}[left=0pt, noitemsep, topsep=0pt]
    \item TimeFloats enables end-to-end floating-point processing, representing both input vectors and weight matrices in the floating-point domain, and producing floating-point outputs.
    
    \item TimeFloats primarily relies on digital building blocks. By using time-domain representations for analog quantities, it avoids complex analog circuits like ADC/DAC, while taking advantages of analog representations to minimize computing complexities through physics-based computing.
    
    \item TimeFloats employs a passive memristor crossbar for training and inference while using CMOS transistors only in the peripherals. This improves area efficiency, enabling the architecture to be folded into two stacks: memristor-based designs implemented at the back-end-of-line (BEOL) and CMOS circuits at the front-end-of-line (FEOL). This results in higher integration density within limited silicon footprint.
    
    \item Although TimeFloats is susceptible to process variability (e.g., voltage and temperature variations), the architecture flexibly incorporates adaptive mechanisms to address these.
\end{itemize}

In Sec. II, we discuss the challenges of train-in-memory to motivate the proposed novelties. Sec. III discusses TimeFloats' architecture and building block. Sec. IV presents simulation results and Sec. V concludes.  

\section{From Compute-in-Memory to Train-in-Memory}
Compute-in-Memory~\cite{9972346,10473702,9180701,giacomini2024towards,9394594} has emerged as a principal technique to significantly enhance the energy efficiency of deep learning at the edge. By utilizing the same memory structure for both storage and computation, the need for data movement of weights, inputs, and product-sums can be minimized. In traditional digital hardware, excessive movement of these operands through limited bandwidth channels forms a key bottleneck for performance scaling and energy reduction. Additionally, compute-in-memory approaches can leverage physics to do away with some of the computations that are necessary for traditional digital systems. For instance, representing element-wise products of weights and inputs as charge or current allows their addition to be performed by Kirchhoff’s law simply over a wire, eliminating the need for a dedicated adder. In this paper, we leverage these leading advantages of compute-in-memory, transforming them for \textit{train-in-memory} so that efficient edge learning can be accomplished.

Our focus on train-in-memory also synergizes with emerging sample-efficient, self-supervised, and real-time learning paradigms. Self-supervised learning, for instance, allows models to make predictions without explicit supervision from human-labeled data. Instead, models learn from the intrinsic structure and patterns present in the data itself. With few-shot learning, predictive models can recognize new objects or classes with only a small number of labeled examples, typically only five or ten. Reinforcement learning allows models to learn by interacting with their environment, while continual learning lets models learn and adapt continuously over time. As data distributions can change over time, and new concepts or classes may need to be added to a model's knowledge, continual learning significantly improves the reliability of machine learning systems in real-world applications. Although the development of these emerging learning paradigms is currently focused mostly on algorithms, train-in-memory can dramatically advance their application prospects by integrating these methods within edge devices such as cell phones, wearables, and AR/VR equipment.

Back-propagation is the dominant algorithm for training neural networks by updating weights to minimize the error between the predicted and true outputs. This process involves two main phases: the \textit{forward pass} and the \textit{backward pass}. In the forward pass, the input vector \(\mathbf{x}_0\) is propagated through the network layer by layer. For each hidden layer, the input is transformed by multiplying it with a weight matrix and applying an activation function. For example, the output of the first hidden layer is given by \(\mathbf{x}_1 = f_1(\mathbf{W}_1 \mathbf{x}_0)\), and the output of the second hidden layer is \(\mathbf{x}_2 = f_2(\mathbf{W}_2 \mathbf{x}_1)\). This process continues until the final output layer, where the output is computed as \(\mathbf{x}_L = f_L(\mathbf{W}_L \mathbf{x}_{L-1})\). The loss \(E\) is then calculated as the difference between the predicted output \(\mathbf{x}_L\) and the true output \(\mathbf{t}\). For regression problems, mean squared error is used: \(E = \frac{1}{2} \|\mathbf{x}_L - \mathbf{t}\|^2\) for loss, while for classification problems, cross-entropy loss is typically used.

In the backward pass, essential for training, the algorithm computes the gradient of the loss with respect to each weight matrix by propagating the error backward through the network. The error at the output layer is calculated first: \(\delta_L = (\mathbf{x}_L - \mathbf{t}) \odot f'_L(\mathbf{W}_L \mathbf{x}_{L-1})\). This error is then propagated backward through the hidden layers using the chain rule: \(\delta_i = (\mathbf{W}_{i+1}^T \delta_{i+1}) \odot f'_i(\mathbf{W}_i \mathbf{x}_{i-1})\). Finally, the weights are updated to minimize the loss. The gradient of the loss with respect to each weight matrix is calculated as \(\frac{\partial E}{\partial \mathbf{W}_i} = \delta_i \mathbf{x}_{i-1}^T\). The weights are then adjusted in the direction that reduces the loss, using learning rate \(\alpha\): \(\mathbf{W}_i \leftarrow \mathbf{W}_i - \alpha \frac{\partial E}{\partial \mathbf{W}_i}\). 

While state-of-the-art memristor crossbars, such as~\cite{8351561}, are suited for \textit{in situ} weight updates and low precision fixed-point arithmetic, a critical advanced functionality needed is the support for floating-point operations to reliably represent a wide range of gradients in the learning process. Therefore, in the subsequent discussion, we demonstrate a novel ``TimeFloats'' microarchitecture that focuses on efficient floating-point multiply-accumulates in the crossbar by pursuing a time-domain representation and primarily relies on digital building blocks to support technology node scalability. 

\begin{figure}[t!]
    \centering
    \includegraphics[width=\linewidth]{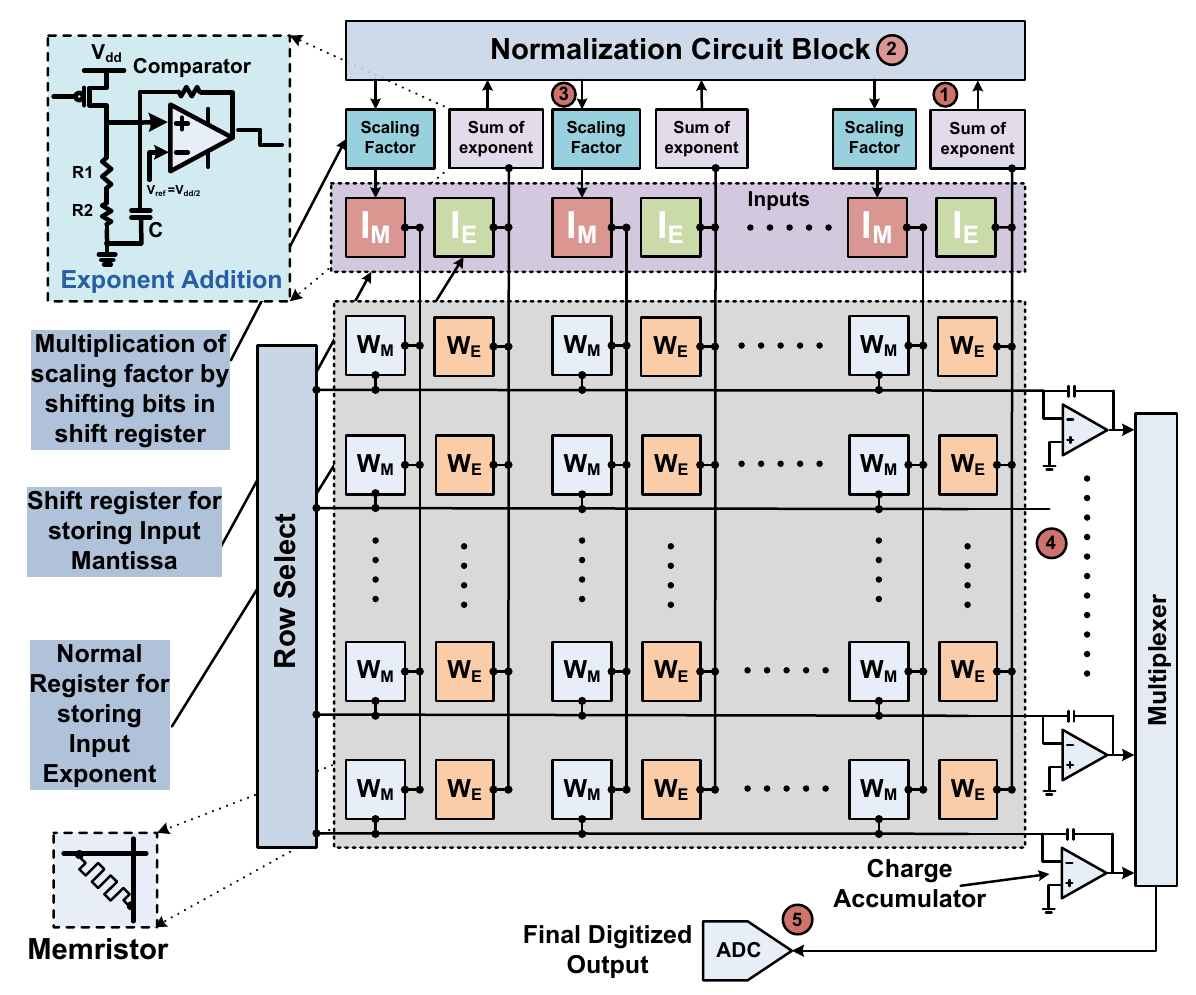}
    \caption{\textbf{Overview of TimeFloats:} TimeFloats utilizes in-memory time-domain processing of floating-point weights and inputs in five steps, as marked in the figure. \textbf{\textit{Step-1:}} Input and weight exponents are added element-wise in a vector-parallel operation using time-domain analog computations. \textbf{\textit{Step-2:}} Exponent sums are normalized across the vector to improve the bit-range of processing in successive steps. \textbf{\textit{Step-3:}} Input mantissa bits are scaled based on the normalizing factor using digital shift operations. \textbf{\textit{Step-4:}} Time-domain multiply-accumulate (MAC) operations are performed between the mantissas of the weights and the inputs. \textbf{\textit{Step-5:}} The net output is digitized and normalized to floating point format.}
    \label{fig:architecture_overview}
\end{figure}

\section{TimeFloats: Floating-point Time-Domain Scalar Products in Memristor Crossbars}

\subsection{Scalar Products in Floating Point Representation}
A floating-point number is typically represented as the product of its sign, mantissa and exponent, where the exponent denotes a power of two. For example: $\text{sign} \times \text{mantissa} \times 2^{\text{exponent}}$. The effective exponent value is obtained by subtracting a reference from the encoded value, resulting in a range from negative to positive (such as -128 to 127). Mantissa's most significant bit (MSB) is consistently set to one and is thus not explicitly stored. Additionally, there exists a sign bit indicating if the floating-point number is negative and positive. Using these representations, the scalar product of input $\mathbf{x}$ and weight $\mathbf{w}$ vectors is computed as:
$\mathbf{y} = \sum_{i=1}^{N} \mathbf{x}_i \cdot \mathbf{w}_i = \sum_{i=1}^{N} (\mathbf{m}_x \cdot \mathbf{m}_w) \times 2^{\mathbf{e}_x + \mathbf{e}_w}$.

\subsection{High-level Overview of TimeFloats}
As evident in previous section, floating-point scalar product computation is more complex than in the fixed-point domain due to the need for both multiplication and renormalization of exponents to ensure proper output format. Fig.~\ref{fig:architecture_overview} provides an overview of the TimeFloats architecture, which performs these operations in-memory in the time-domain. TimeFloats processes the applied floating point input vector against the stored weight matrix in five steps (as also identified in Fig.~\ref{fig:architecture_overview}): \darkgreycircled{1} element-wise exponent addition, \darkgreycircled{2} across column normalization of added exponents, \darkgreycircled{3} element-wise input mantissa scaling based on the normalization factor, \darkgreycircled{4} fixed-point scalar products between normalized input mantissa vector and weight vector, and \darkgreycircled{5} digitization and product-sum's normalization to floating-point format. 

Correspondingly, the architecture consists of five key building blocks: (i) a memristor crossbar at the center, with each cell containing two memristors storing 4-bit mantissa and exponent information for the matrix weights; (ii) an exponent addition block at the top, which stores input vectors' exponent bits in registers and uses analog circuits for time-domain addition to the exponents of the weights stored in the crossbar; (iii) a normalization circuit block, which normalizes exponent bits across columns and identifies the largest exponents among the inputs; (iv) a shift-scaling block, which scales all input mantissas based on the normalization to maximize operating range; and (v) peripherals at the bottom for digitizing and routing column-computed outputs.

An additional important implementation advantage of TimeFloats is to allow vertical folding of the architecture for physical design. Notably, TimeFloats uses only a passive memristor crossbar (i.e, without access transistors) so that the crossbar can be implemented at the back end of the line (BEOL) process steps, while all peripherals can be implemented underneath using front end of the line (FEOL) process steps. This folding of components into two layers minimizes the area required for the design, thus optimizing the silicon footprint to implement multiple crossbars even in limited space. Subsequently, we discuss TimeFloats' operation with 8-bit floating point precision--4-bit mantissa and 4-bit exponent. While the architecture can be flexibly modified for other floating point precisions, training methods for deep neural networks now support learning with such limited bandwidth. 

\begin{figure}[t!]
    \centering
    \includegraphics[width=\linewidth]{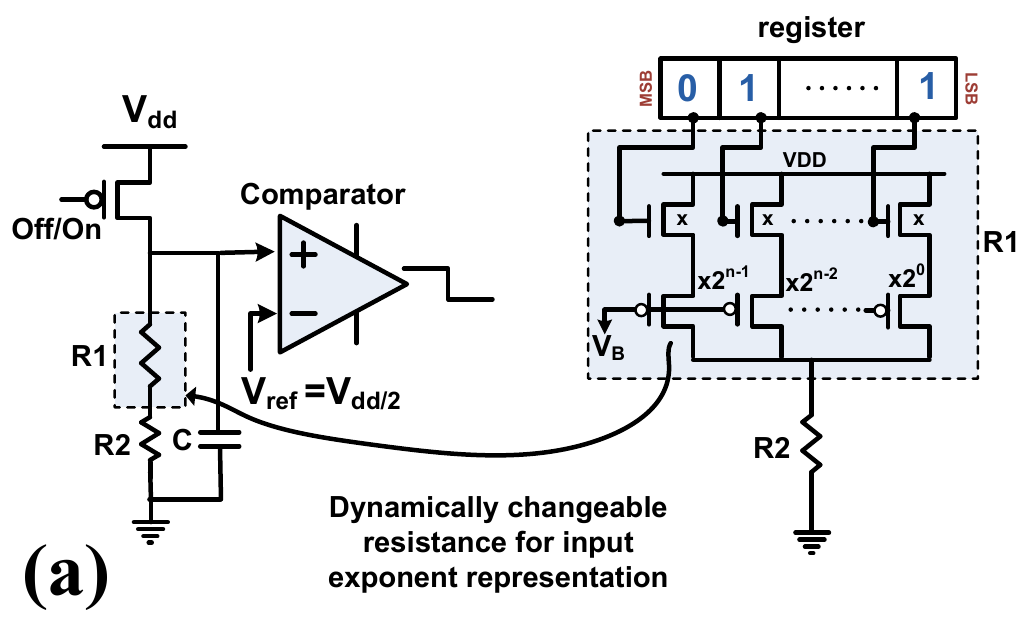}
    \includegraphics[width=0.8\linewidth]{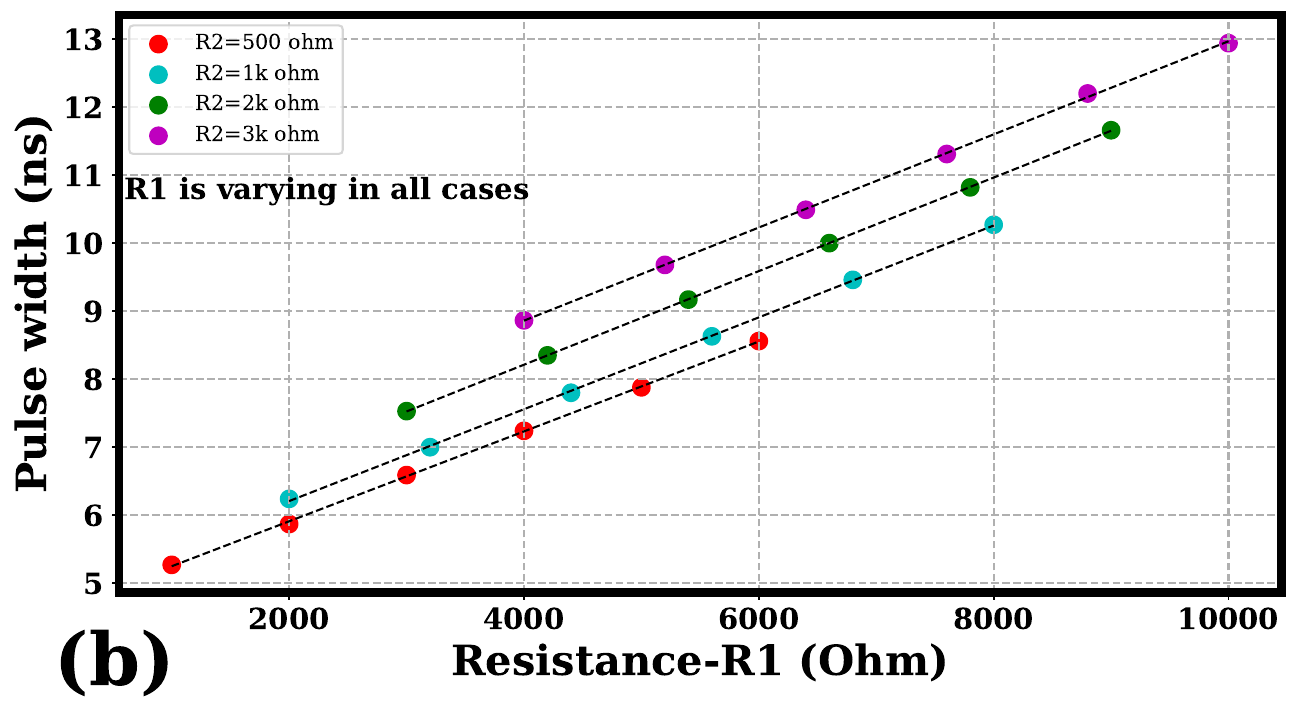}
    \caption{\textbf{Exponent Adder:} \textbf{(a)} We use the RC path discharge for element-wise exponent summation and time-domain conversion of the output for subsequent processing. \textbf{(b)} Linearity of exponent summations at varying input/weight mantissa bit combinations.}
    \label{fig:exponent_addition_circuit}
\end{figure}

\subsection{TimeFloats' Building Blocks}

\noindent\textbf{(i) Exponent Adder:} Fig. \ref{fig:exponent_addition_circuit}(a) shows the exponent adders in TimeFloats. The exponent adder performs element-wise addition of input and weight exponents and produces the resultant output in the time domain for subsequent processing. The delay of the $RC$-path is utilized for summation and conversion to the time domain. For this, digital bits for input exponent are first converted to equivalent resistance ($I_E$) using the circuits shown in Fig.~\ref{fig:exponent_addition_circuit}(a) which is adjoined to the crossbar along the column bitlines connected with memristors storing weight exponents. The corresponding memristive weight in the row ($W_E$) is selected by grounding its row terminal whereas the exponent memristors in the other rows are left floating. The series connected resistance thereby follows the the summation of input and weight exponents, i.e., $W_E + I_E$.  The top-terminal of the series-connected resistances is precharged to VDD and subsequently let discharge. A clocked comparator converts the discharge transient to time-pulse which follows the summation of input and weight exponents.  

Fig.~\ref{fig:exponent_addition_circuit}(b) demonstrates the linearity of the exponent sums considering various input and weight-bit mantissa combinations, indicating sufficient linearity. Notably, while the analog processing of exponent summation is susceptible to process variability, several calibration knobs can be inserted in the above. First, the bias voltage $V_B$ in Fig.~\ref{fig:exponent_addition_circuit}(a) can be tuned to collectively counter variability of all columns in a crossbar. Complementary to this, column-to-column variability can be addressed by connecting parallel calibration memristors to R2 in the circuit and programming them. The process variability of exponent memristors ($W_E$) in the crossbar can be corrected through dynamic tuning of the resistance states itself. For example, calibration circuits, as presented in prior work~\cite{9524794}, can implement program-read-tune cycles to correct the variability.

\begin{figure}[t!]
    \centering
    \includegraphics[width=\linewidth]{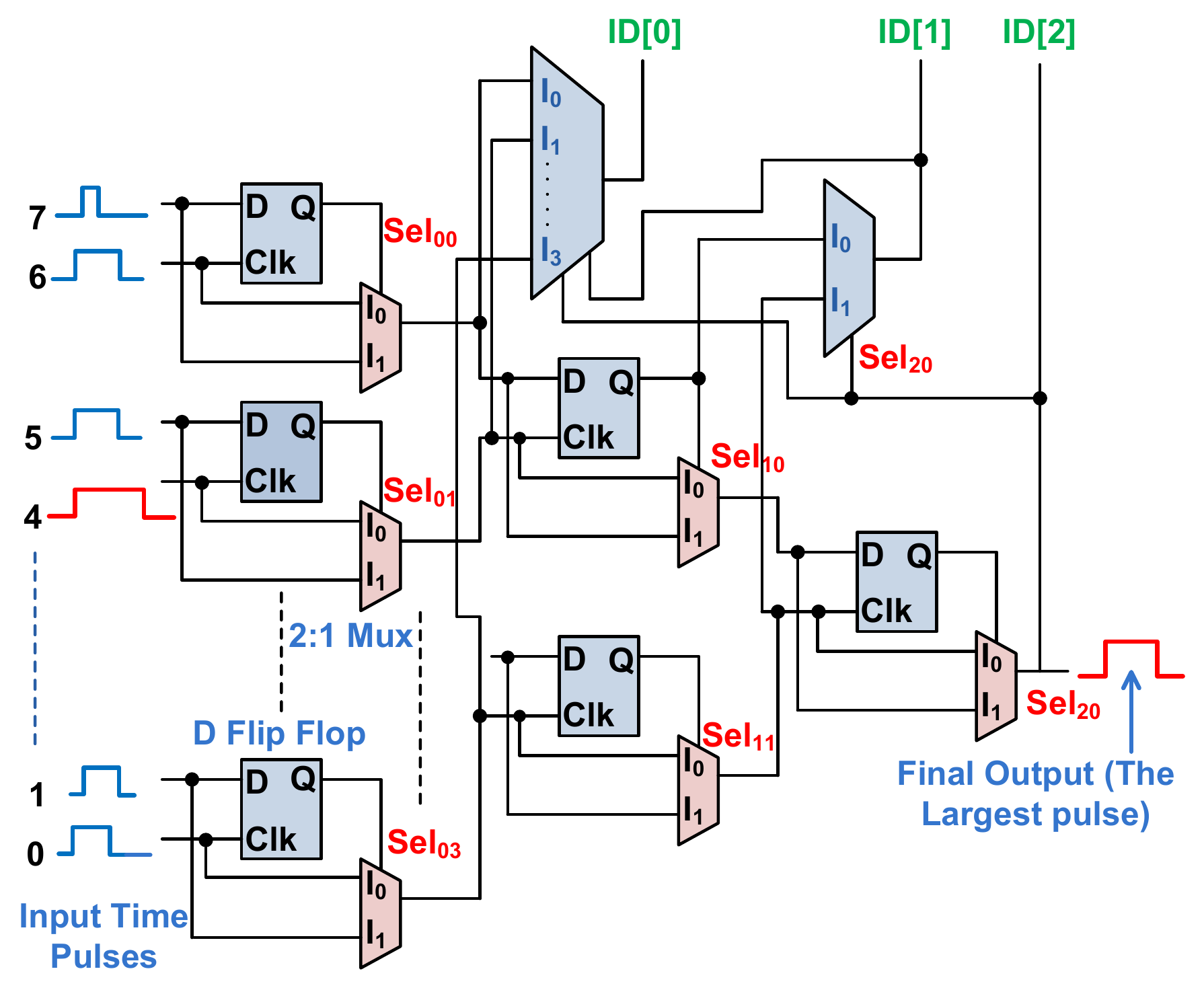}
    \caption{\textbf{Largest Exponent Detector:} After element-wise exponent additions, the largest summed exponent is identified for normalization using a tree-like structure of D Flip-Flops and 2:1 multiplexers. The argument of largest sum is determined using select pin outputs.}
    \label{fig:Exp_search}
\end{figure}

\vspace{4pt}
\noindent\textbf{(ii) Largest Exponent Detector:} After element-wise exponent additions, the largest summed exponent is identified for normalization. Fig.~\ref{fig:Exp_search} shows a digital circuit for detecting the largest resultant exponent using a tree-like structure of D Flip-Flops and 2:1 multiplexers (mux). When two resultant exponents, represented by time pulses, are applied to the D Flip-Flop's input and clock pins, the output signal at the clock's negative edge indicates whether the input pulse is larger than the clock pulse. A longer input pulse results in a high output, while a shorter pulse yields a low output. By feeding these signals into a 2:1 mux, with the select pin connected to the D Flip-Flop output, the mux delivers the longest time pulse. Replicating this D Flip-Flop and mux block in a tree-like structure ensures that the longest pulse emerges at the end. To determine the ID of the largest pulse, we use select pin outputs from various stages of the detection circuit. For 8 signals, 3 bits are required to represent each ID. Fig.~\ref{fig:Exp_search} shows a series of muxes designed for this purpose, providing the ID bits of the largest time pulse.

\begin{figure}[t!]
    \centering
    \includegraphics[width=\linewidth]{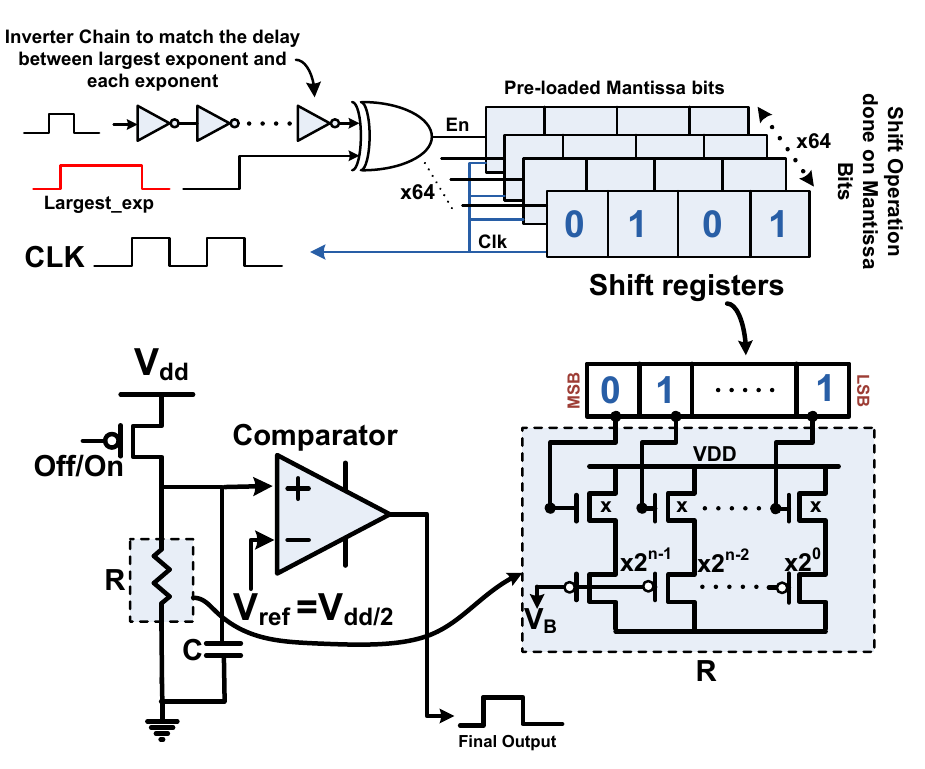}
    \caption{\textbf{Exponent Normalization:} After identifying the largest exponent, it is subtracted from the other exponent terms in time-domain itself. Based on the resultant, other mantissa terms are scaled. THe scaled mantissa is then converted to time pulse for subsequent processing.}
    \label{fig:Exp_normalize}
\end{figure}

\begin{figure}[t!]
    \centering
    \includegraphics[width=\linewidth]{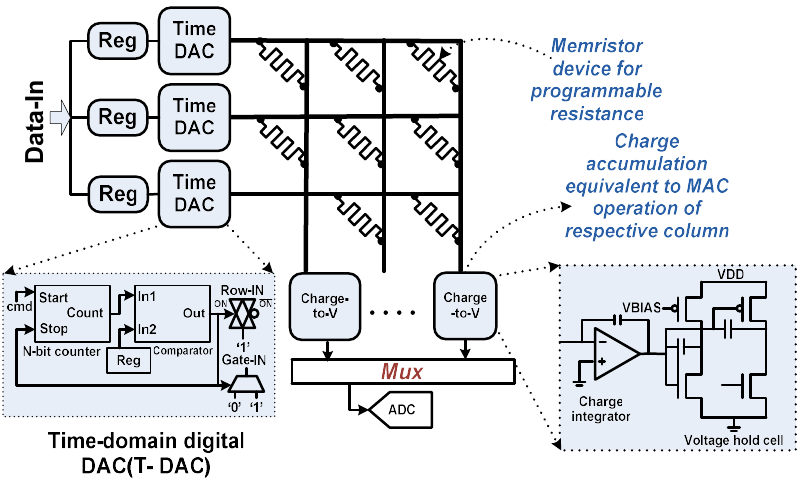}
    \caption{\textbf{Crossbar circuit for MAC operation:} Memristor-based crossbar with charge accumulator for fixed point operations on scaled mantissas. }
    \label{fig:crossbar_explained}
\end{figure}

\begin{figure}[t!]
    \centering
    \includegraphics[width=\linewidth]{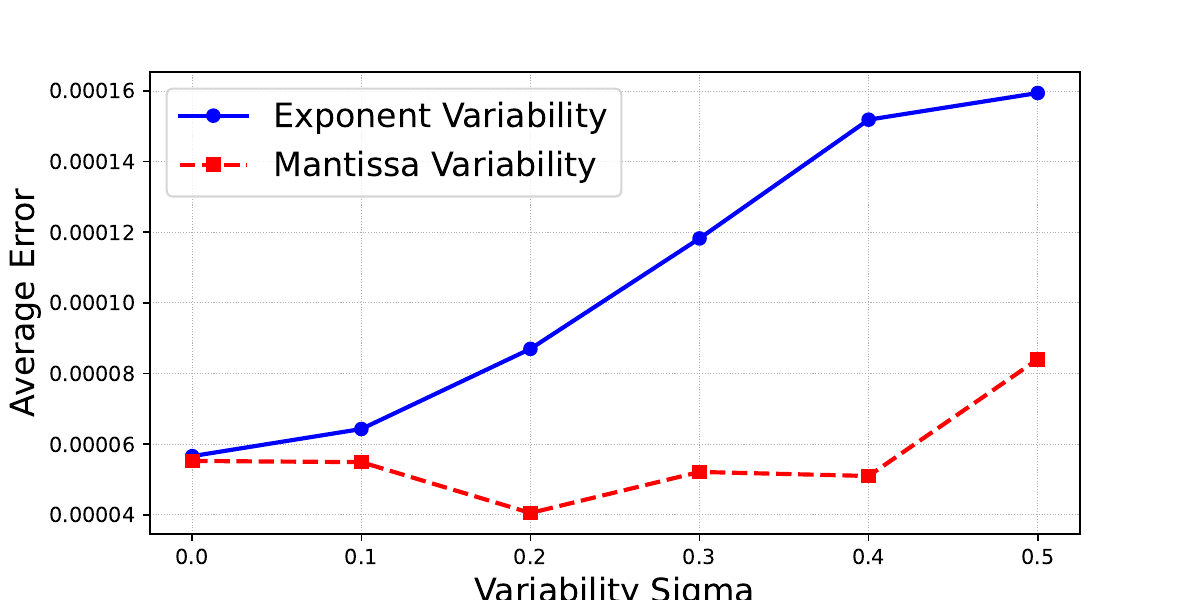}
    \caption{\textbf{Impact of process variability:} Separately considering process variability's impact on exponent vs. mantissa computations, a higher sensitivity of error to exponent variability is demonstrated.}
\end{figure}

\begin{table*}[t!]
\footnotesize
\setlength{\tabcolsep}{6pt}
\renewcommand{\arraystretch}{1}
\centering
\captionsetup{font=small}
\caption{\textbf{Breakdown of TimeFloats' Operational Energy for 64-element 8-bit Input-Weight Scalar Product in Floating Point}}
\begin{tabularx}{\linewidth}{cccc}
\hline
 & \textbf{Floating Point MAC Operation Steps} & \textbf{TimeFloat's Module} & \textbf{Energy} \\ \hline
\darkgreycircled{1} & Element-wise input-weight exponent addition: $W_E^i + I_E^i$ & Mixed-signal exponent adder in Fig. 3 & 1.28 pJ \\ 
\darkgreycircled{2} & Searching largest added-exponent: $E_\text{max} = \text{argmax}(W_E^i + I_E^i)$ & D-F/F and MUX Tree in Fig. 4 & 3.25 pJ \\ 
\darkgreycircled{3} & Element-wise input mantissa scaling: $I_\text{M,scaled}^i = I_M^i/2^{(E_\text{max} - W_E^i - I_E^i)}$ & Time-domain subtraction and right shift in Fig. 5 & 23 fJ \\
\darkgreycircled{4} & Scalar product of scaled input-weight mantissa vectors: $P_\text{sum} = \sum_i I_\text{M,scaled}^i \times W_\text{M}^i$ & Memristor crossbar in Fig. 6 & 1.23 pJ \\ 
\darkgreycircled{5} & Product-sum digitization: $D_\text{OUT} = \text{ADC}(P_\text{sum})$ and reformatting & 4-bit Analog-to-Digital Converter (ADC) & 21 fJ \\ \hline
\end{tabularx}
\end{table*}

\vspace{4pt}
\noindent\textbf{(iii) Time-Domain Exponent Normalization and Mantissa Scaling:} 
Subsequent to identifying the largest exponent ($E_\text{max}$), we utilize its time-domain representation to normalize the other exponents, thereby increasing the dynamic range of processing and eliminating redundant computations under prescribed quantization. Specifically, exponent terms in the input vector ($I_E^i$) are scaled by subtracting them from the largest exponent, i.e., $E_\text{max} - I_E^i$, directly in the time domain. Following this, their mantissa bits are scaled as $I_{\text{M,scaled}}^i = \frac{I_M^i}{2^{(E_\text{max} - W_E^i - I_E^i)}}$. Consequently, if $2^{(E_\text{max} - W_E^i - I_E^i)}$ exceeds the mantissa bit width, denoted as $2^{N_M}$ (where $N_M$ is the number of bits for the mantissa), the corresponding mantissa term is scaled to zero. This approach thus not only improves the dynamic range in our floating-point MAC design but also eliminates redundant computations, enhancing sparsity.

To implement this processing in the time domain, we present the TimeFloats' scheme, illustrated in Fig. 5. Here, the time-domain converted exponents $I_E$ are passed through an inverter chain that matches the delay of the exponent detector circuit depicted in Fig. 4. Subsequently, the XOR gate processes the delay input pulse and the largest exponent pulse to compute their difference. Based on this difference, the shift register, in which the mantissa bits are initially loaded, shifts them by one place with each clock cycle. This shifting effectively performs the division by $2^{(E_\text{max} - W_E^i - I_E^i)}$. Hence, if the difference between the pulses exceeds $2^N$, the mantissa bits are zeroed due to the large scaling factor.

The remaining mantissa bits are then loaded into a register-conversion circuit, as depicted in the lower half of Fig. 5. In this step, we again utilize our RC coding approach to convert the mantissa bits into a time pulse, which is then used for subsequent time-domain processing within the crossbar.

\vspace{4pt}
\noindent\textbf{(iv) Fixed-Point Input-Weight Mantissa Scalar Product:} Subsequently, a fixed-point scalar product of input-weight mantissa vectors is performed as shown in Fig.~\ref{fig:crossbar_explained}. TimeFloats' memristor crossbar array consists of a grid where each intersection contains a memristor with conductance (\(G_{ij}\)) representing a matrix element. Unlike other designs\cite{8551115}, we utilize a fully-passive crossbar to support vertical folding atop transistor-based peripheral circuits. Inputs are applied in the time domain using a digital-to-pulse converter (T-DAC) composed entirely of digital components. Crossbars operate on time-encoded input signals against stored weights in the charge domain, with charge accumulated along a column through an integrator circuit. The potential developed across the integrating capacitor \(C_{\text{int}}\) linearly follows \(\sum_i T_i \times g_{ij}\), where \(T_i\) is the pulse width of the encoded input vector element at row \(i\) and \(g_{ij}\) is the programmed conductance at the \(i\)th row and \(j\)th column. The integrated charge can be briefly held using a voltage hold cell. We used TiO\textsubscript{2}-based memristors with resistance with range 0.1 M\(\Omega\) -- 1 M\(\Omega\).

\vspace{4pt}
\noindent\textbf{(v) Digitization and Reformatting:} At the end of the memristor crossbar architecture, each column is connected to an operational amplifier (op-amp) with a feedback capacitor acting as a charge accumulator. The voltage output from the capacitor is held in a transistor-based voltage hold cell, which temporarily stores the analog signal. This stored voltage is then digitized for further processing and storage. To minimize the use of ADCs for digitization, a single ADC is shared among all columns through a MUX. The digitized data is subsequently reformatted to floating-point format for routing and storage.

\subsection{Design Space Guidance against Process Variability}
Analog operations, despite time-domain representations in TimeFloats, invariably introduce accuracy degradation due to process variability. For instance, converting digital exponential bits to time-domain representations through the resistive circuit shown in Fig. 2 is affected by the variability in transistors and memristors, leading to inaccurate bits-to-time conversion. While this variability can be mitigated by exploiting the non-volatile programmability of memristors in the circuit for calibration, it cannot be entirely eliminated, resulting in residual variability. Similarly, when converting product-sum outputs to digital form, the process variability in analog components like multiplexers and ADCs degrades accuracy. Although additional design resources could be allocated to counteract these variabilities, edge devices are inherently constrained in terms of resources like area, power, and time, limiting the extent to which these variabilities can be mitigated. Therefore, in Fig. 7, we analyze the sensitivity of process variability to exponent and mantissa computations, respectively, to guide designers in resource allocation. In the figure, we model process variability using a Gaussian distribution, where ideal computation output $C$ is perturbed as $C \rightarrow C\times(1+\mathcal{N}(0,\sigma))$, where $\mathcal{N}(0,\sigma)$ is the representative net process variability impact. Considering such parametric model, we assessed the impact on accuracy through 100 Monte Carlo simulations across varying levels of process variability ($\sigma$). The results in the figure show that mantissa computations are relatively stable across the considered variability range, whereas exponent computations are much more sensitive. Therefore, under limited resources, they can be focused more on countering process variability in exponent computations to mitigate their higher sensitivity.

\section{Simulation Results and Discussions}

\subsection{Simulation Methodology}

\begin{table*}
\footnotesize
    \setlength{\tabcolsep}{8pt}
    \renewcommand{\arraystretch}{1}
    \centering 
    \captionsetup{font=small}
    \caption{\textbf{Comparison between our approach and state-of-the-art for multiply-accumulate (MAC) processing}}
    \label{tab:comparison}
    \begin{tabularx}{\linewidth}{ccccccccc}
        \hline
        \textbf{Metric} & \textbf{Ours} & \textbf{\cite{10067527}} & \textbf{\cite{10067260}} & \textbf{\cite{9731681}} & \textbf{ \cite{10454313}} & \textbf{\cite{10265269}} & \textbf{\cite{9365984}}\\
        \hline
        Technology & \textbf{15nm} & 22nm & 28nm & 28nm & 28nm & 22nm & 28nm \\
        MAC Operation Domain  & \textbf{Time } & Hybrid  & Digital & Time & Hybrid & Hybrid & Analog \\
        Input Precision & \textbf{FP8} & BF16 & BF16/INT8 & INT8/INT4 & BF16/INT8 & BF16 & INT8/INT4 \\
        Weight Precision & \textbf{FP8} & BF16 & BF16/INT8 & INT8/INT4 & BF16/INT8 & BF16 & INT8/INT4\\
        Memory type & \textbf{Memristor} & SRAM & SRAM & SRAM & SRAM & SRAM & SRAM\\
        Energy Efficiency (TOPS/W) & 22.1 & 16.22-17.59  & 19.5-44 & 21.10-27.75 & 22.78-50.53 & 72.12 & 15.02-22.75\\              
        \hline
    \end{tabularx}
\end{table*}

In previous discussions, we outlined how TimeFloats leverages both digital and analog representations to maximize efficiency. Digital building blocks enhance integrability with other computing modules on a chip, while time-based analog representations minimize computational complexity through physics-based computations, particularly for memristive computations where the use of time pulses enables charge accumulation for product terms, which are added using a charge integrator, obviating a dedicated adder circuit. Integration of both analog and digital operations within the same microarchitecture thus requires special considerations on simulation methodologies. The digital modules were designed using SystemVerilog and synthesized with Synopsys Design Compiler, targeting the 15nm NCSU predictive technology library. The analog modules, including circuits for exponent addition, dynamically adjustable resistance, and a memristor crossbar, were designed and simulated in HSPICE with the 15nm predictive technology. The use of SystemVerilog allows us to rapidly synthesize the peripherals and digital building blocks by enforcing time constraints. Meanwhile, detailed SPICE simulations of the inner crossbar cells provides insights into the impact of process variability and the non-idealities of analog computations on the system's overall behavior.

\subsection{Operating Energy Characterization}
Table I analyzes the energy breakdown of TimeFloats operations for various tasks in floating-point scalar product computations. We consider a 64-element processing on a 64×128 memristor crossbar. Each cell stores 4-bit information using two memristors--one 4-bit memristor storage for exponent bits and another 4-bit memristor storage for mantissa bits. The first operation, element-wise input-weight scaling, uses the mixed-signal exponent adder (Fig. 3) and consumes 1.28 pJ. Finding the largest exponent, implemented with a synthesized flip-flop in SystemVerilog, consumes 3.25 pJ. Element-wise mantissa scaling, performed via right shifting and time-domain difference by XOR, requires 23 fJ. Scalar products, computed using memristors with a maximum pulse width of 15 ns for 4-bit input application, consume 1.32 pJ. Finally, sum digitization using an SAR ADC consumes 2.421 pJ for 4-bit conversions. The total energy for 64-element processing (i.e., 128 floating-point MAC operations) is 5.8 pJ, yielding an efficiency of 22.1 TOPS/W.

\section{Conclusion}
In Table II, our proposed design outperforms competitive work in terms of energy efficiency, achieving 22.1 TOPS/W for full end-to-end floating-point computations. This efficiency stems from time-domain operations, which simplify the process by utilizing analog building blocks while still leveraging analog representations. Notably, our design only requires one ADC for full row processing. Further opportunities for optimization exist. In future work, we plan to exploit the exponent detector for predictive early termination in crossbar processing, which will significantly reduce overall power consumption. Additionally, custom design of the largest exponent detector could lower its power usage, which is currently the largest contributor to power in our design. 


\vspace{5pt}
\noindent\textbf{Acknowledgement: }This work was supported by NSF Future of Semiconductors (FuSe) Award \#2329096 and Department of Energy Office of Science Award \#DE-SC0023715.

\bibliographystyle{IEEEtran}
\bibliography{ref.bib}

\begin{thebibliography}{10}
\providecommand{\url}[1]{#1}
\csname url@samestyle\endcsname
\providecommand{\newblock}{\relax}
\providecommand{\bibinfo}[2]{#2}
\providecommand{\BIBentrySTDinterwordspacing}{\spaceskip=0pt\relax}
\providecommand{\BIBentryALTinterwordstretchfactor}{4}
\providecommand{\BIBentryALTinterwordspacing}{\spaceskip=\fontdimen2\font plus
\BIBentryALTinterwordstretchfactor\fontdimen3\font minus \fontdimen4\font\relax}
\providecommand{\BIBforeignlanguage}[2]{{%
\expandafter\ifx\csname l@#1\endcsname\relax
\typeout{** WARNING: IEEEtran.bst: No hyphenation pattern has been}%
\typeout{** loaded for the language `#1'. Using the pattern for}%
\typeout{** the default language instead.}%
\else
\language=\csname l@#1\endcsname
\fi
#2}}
\providecommand{\BIBdecl}{\relax}
\BIBdecl

\bibitem{gupta2015deep}
S.~Gupta, A.~Agrawal, K.~Gopalakrishnan, and P.~Narayanan, ``Deep learning with limited numerical precision,'' in \emph{International conference on machine learning}.\hskip 1em plus 0.5em minus 0.4em\relax PMLR, 2015, pp. 1737--1746.

\bibitem{9972346}
P.~Shukla, S.~Nasrin, N.~Darabi, W.~Gomes, and A.~R. Trivedi, ``Mc-cim: Compute-in-memory with monte-carlo dropouts for bayesian edge intelligence,'' \emph{IEEE Transactions on Circuits and Systems I: Regular Papers}, vol.~70, no.~2, pp. 884--896, 2023.

\bibitem{10473702}
N.~Darabi, M.~B. Hashem, H.~Pan, A.~Cetin, W.~Gomes, and A.~R. Trivedi, ``Adc/dac-free analog acceleration of deep neural networks with frequency transformation,'' \emph{IEEE Transactions on Very Large Scale Integration (VLSI) Systems}, vol.~32, no.~6, pp. 991--1003, 2024.

\bibitem{9180701}
P.~Shukla, A.~Shylendra, T.~Tulabandhula, and A.~R. Trivedi, ``Mc2ram: Markov chain monte carlo sampling in sram for fast bayesian inference,'' in \emph{2020 IEEE International Symposium on Circuits and Systems (ISCAS)}, 2020, pp. 1--5.

\bibitem{giacomini2024towards}
D.~Giacomini, M.~B. Hashem, J.~Suarez, and A.~R. Trivedi, ``Towards model-size agnostic, compute-free, memorization-based inference of deep learning,'' in \emph{2024 37th International Conference on VLSI Design and 2024 23rd International Conference on Embedded Systems (VLSID)}.\hskip 1em plus 0.5em minus 0.4em\relax IEEE, 2024, pp. 180--185.

\bibitem{9394594}
S.~Nasrin, D.~Badawi, A.~E. Cetin, W.~Gomes, and A.~R. Trivedi, ``Mf-net: Compute-in-memory sram for multibit precision inference using memory-immersed data conversion and multiplication-free operators,'' \emph{IEEE Transactions on Circuits and Systems I: Regular Papers}, vol.~68, no.~5, pp. 1966--1978, 2021.

\bibitem{8351561}
A.~Haj-Ali, R.~Ben-Hur, N.~Wald, and S.~Kvatinsky, ``Efficient algorithms for in-memory fixed point multiplication using magic,'' in \emph{2018 IEEE International Symposium on Circuits and Systems (ISCAS)}, 2018.

\bibitem{9524794}
J.~Fu, Z.~Liao, and J.~Wang, ``Cycle-to-cycle variation enabled energy efficient privacy preserving technology in ann,'' in \emph{2020 IEEE 33rd International System-on-Chip Conference (SOCC)}, 2020, pp. 66--71.

\bibitem{8551115}
N.~Athreyas, Q.~Xia, D.~Gupta, W.~Song, A.~Mathew, J.~J. Yang, B.~Perot, and J.~Gupta, ``Memristor-cmos analog co-processor for acceleration of high performance computing applications,'' in \emph{International Conference on Current Trends towards Converging Technologies}, 2018, pp. 1--7.

\bibitem{10067527}
P.-C. Wu, J.-W. Su, L.-Y. Hong, J.-S. Ren, C.-H. Chien, H.-Y. Chen, C.-E. Ke, H.-M. Hsiao, S.-H. Li, S.-S. Sheu, W.-C. Lo, S.-C. Chang, C.-C. Lo, R.-S. Liu, C.-C. Hsieh, K.-T. Tang, and M.-F. Chang, ``A 22nm 832kb hybrid-domain floating-point sram in-memory-compute macro with 16.2-70.2tflops/w for high-accuracy ai-edge devices,'' in \emph{2023 IEEE International Solid-State Circuits Conference (ISSCC)}, 2023.

\bibitem{10067260}
A.~Guo, X.~Si, X.~Chen, F.~Dong, X.~Pu, D.~Li, Y.~Zhou, L.~Ren, Y.~Xue, X.~Dong, H.~Gao, Y.~Zhang, J.~Zhang, Y.~Kong, T.~Xiong, B.~Wang, H.~Cai, W.~Shan, and J.~Yang, ``A 28nm 64-kb 31.6-tflops/w digital-domain floating-point-computing-unit and double-bit 6t-sram computing-in-memory macro for floating-point cnns,'' in \emph{2023 IEEE International Solid-State Circuits Conference (ISSCC)}, 2023.

\bibitem{9731681}
P.-C. Wu, J.-W. Su, Y.-L. Chung, L.-Y. Hong, J.-S. Ren, F.-C. Chang, Y.~Wu, H.-Y. Chen, C.-H. Lin, H.-M. Hsiao, S.-H. Li, S.-S. Sheu, S.-C. Chang, W.-C. Lo, C.-C. Lo, R.-S. Liu, C.-C. Hsieh, K.-T. Tang, C.-I. Wu, and M.-F. Chang, ``A 28nm 1mb time-domain computing-in-memory 6t-sram macro with a 6.6ns latency, 1241gops and 37.01tops/w for 8b-mac operations for edge-ai devices,'' in \emph{2022 IEEE International Solid-State Circuits Conference (ISSCC)}, vol.~65, 2022, pp. 1--3.

\bibitem{10454313}
Y.~Yuan, Y.~Yang, X.~Wang, X.~Li, C.~Ma, Q.~Chen, M.~Tang, X.~Wei, Z.~Hou, J.~Zhu, H.~Wu, Q.~Ren, G.~Xing, P.-I. Mak, and F.~Zhang, ``34.6 a 28nm 72.12tflops/w hybrid-domain outer-product based floating-point sram computing-in-memory macro with logarithm bit-width residual adc,'' in \emph{2024 IEEE International Solid-State Circuits Conference (ISSCC)}, vol.~67, 2024, pp. 576--578.

\bibitem{10265269}
P.-C. Wu, J.-W. Su, L.-Y. Hong, J.-S. Ren, C.-H. Chien, H.-Y. Chen, C.-E. Ke, H.-M. Hsiao, S.-H. Li, S.-S. Sheu, W.-C. Lo, S.-C. Chang, C.-C. Lo, R.-S. Liu, C.-C. Hsieh, K.-T. Tang, and M.-F. Chang, ``A floating-point 6t sram in-memory-compute macro using hybrid-domain structure for advanced ai edge chips,'' \emph{IEEE Journal of Solid-State Circuits}, vol.~59, no.~1, pp. 196--207, 2024.

\bibitem{9365984}
J.-W. Su, Y.-C. Chou, R.~Liu, T.-W. Liu, P.-J. Lu, P.-C. Wu, Y.-L. Chung, L.-Y. Hung, J.-S. Ren, T.~Pan, S.-H. Li, S.-C. Chang, S.-S. Sheu, W.-C. Lo, C.-I. Wu, X.~Si, C.-C. Lo, R.-S. Liu, C.-C. Hsieh, K.-T. Tang, and M.-F. Chang, ``16.3 a 28nm 384kb 6t-sram computation-in-memory macro with 8b precision for ai edge chips,'' in \emph{2021 IEEE International Solid-State Circuits Conference (ISSCC)}, vol.~64, 2021, pp. 250--252.

\end{thebibliography}
\end{document}